\documentclass{optica-article}

\journal{opticajournal} 
\articletype{Preprint}

\usepackage{lineno}

\begin{document}

\title{Controllable nonlinear propagation of partially incoherent Airy beams}
\noindent
 \textbf{Kaijian Chen,\authormark{1} Peiyu Zhang,\authormark{1} Nana Liu,\authormark{1} Liu Tan,\authormark{1} Peilong Hong,\authormark{2,3,*} Bingsuo Zou,\authormark{4} Jingjun Xu,\authormark{3} and Yi Liang\authormark{1,3,5,*}}

\address{\authormark{1}Guangxi Key Lab for Relativistic Astrophysics, Center on Nanoenergy Research, School of Physical Science and Technology, Guangxi University, Nanning, Guangxi 530004, China\\
\authormark{2}School of Optoelectronic Science and Engineering, University of Electronic Science and Technology of China (UESTC), Chengdu 610054, China\\
\authormark{3}The MOE Key Laboratory of Weak-Light Nonlinear Photonics, TEDA Applied Physics Institute and School of Physics, Nankai University, Tianjin 300457, China\\
\authormark{4}School of Physical Science and Technology and School of Resources, Environment and Materials, Key Laboratory of new Processing Technology for Nonferrous Metals and Materials, Guangxi University, Nanning 530004, China\\
\authormark{5}State Key Laboratory of Featured Metal Materials and Life-cycle Safety for Composite Structures, Nanning 530004, China}

\email{\authormark{*}plhong@uestc.edu.cn} 
\email{\authormark{*}liangyi@gxu.edu.cn}

\begin{abstract*} 
The self-accelerating beams such as the Airy beam show great potentials in many applications including optical manipulation, imaging and communication. However, their superior features during linear propagation could be easily corrupted by optical nonlinearity or spatial incoherence individually. Here we investigate how the interaction of spatial incoherence and nonlinear propagation affect the beam quality of Airy beam, and find that the two destroying factors can in fact balance each other. Our results show that the influence of coherence and nonlinearity on the propagation of partially incoherent Airy beams (PIABs) can be formulated as two exponential functions that have factors of opposite signs. With appropriate spatial coherence length, the PIABs not only resist the corruption of beam profile caused by self-focusing nonlinearity, but also exhibits less anomalous diffraction caused by the self-defocusing nonlinearity. Our work provides deep insight into how to maintain the beam quality of self-accelerating Airy beams by exploiting the interaction between partially incoherence and optical nonlinearity. Our results may bring about new possibilities for optimizing partially incoherent structured field and developing related applications such as optical communication, incoherent imaging and optical manipulations.

\end{abstract*}

\section{Introduction}
Partially incoherent structured light has attracted a lot of interests recently\cite{RN7,RN5,RN10,RN3}, since the control of coherence length provides new features into the structured light that typically originates from optical interference. Interestingly, partially incoherent structured light can present better robustness in turbulent atmosphere and random media\cite{RN6,RN11}, leading to superior performance in many intriguing applications such as optical communication\cite{RN6}, incoherent imaging\cite{RN8}, optical manipulations\cite{RN4}, optical nonlinear solitons or lattices\cite{RN18} and so on.

Airy beams have been widely employed in optical manipulations\cite{RN9}, plasmons\cite{RN19} and optical image transmissions\cite{RN14,RN16} due to their intriguing and counterintuitive properties\cite{RN22,RN23}: self-accelerating, non-diffracting and self-healing. As one of the typical self-accelerating beams, Airy beams in the linear systems have been studied fully in previous works\cite{RN12,RN17}. Beyond  the linear systems, Yi Hu \textit{et al} reported the diffraction behavior of Airy beams in photorefractive crystals\cite{RN21}, which had led to the discussion of the nonlinear propagation properties of Airy beams, such as self-trapped waves and spatial solitons\cite{RN20,RN18}. Then Noémi Wiersma \textit{et al}\cite{RN13} explored the spatial solitons generated by one-dimension Airy beams passing through photorefractive crystals and discussed the influence of an off-shooting soliton on self-focusing nonlinearity. In fact, These results indicate that a strong self-focusing nonlinearity corrupts the beam structure and self-accelerating propagation.Thus, it is challenging yet meaningful to optimize the self-accelerating beams for maintaining their excellent features during nonlinear propagation.

However, to the best of our konwledge, in compare with the thorough investigation of nonlinear dynamics of fully coherent self-accelerating beams, self-accelerating beams with a low spatial coherence, i.e. partially incoherent self-accelerating beams, have not been investigated in nonlinear media. By appropriately controlling the coherence length, R. Martínez-Herrero \textit{et al}\cite{RN2} theoretically demonstrated that partially incoherent Airy beams can still maintain their own unique propagation properties by using cross-spectral-density. In addition, Z. Pang \textit{et al}\cite{RN31} investigated the partially coherent  multi-Airy beams, which provides new understanding for controllable linear propagation of partially coherent accelerating beams. But, this interesting work only focuses on the physical properties in the linear case without discussing nonlinear propagation. Nonetheless, similar to nonlinear propagation, partially coherence was usually  thought corrupting the beam quality of self-focusing beam\cite{RN24,RN2}. Here, by thoroughly studying the nonlinear propagation of partially incoherent Airy beam, we found that, surprisingly, under proper condition the partially coherence could resist the corruption of beam quality caused by nonlinear  propagation. To demonstrate such an interesting results, we first study the propagation of partially incoherent Airy beams driven by nonlinearity experimentally. The results show that, compared with the nonlinear propagation of fully-coherent Airy beams, PIABs can reduce the nonlinear effect and keep their shape better. In other words, we demonstrate that a partially incoherent self-accelerating beam can maintain its shape easier in a nonlinear media and exhibit a good self-accelerating performance in subsequent propagation. For self-focusing case, within the appropriate range, with more spatial incoherence and the less nonlinearity, the beams present better self-accelerating and shape-preserving propagation. Counter to that, for self-defocusing case, the beams show a better performance until reaching a saturable state when self-defocusing nonlinearity is too large. Further theoretical analysis shows that the nonlinearity and spatial coherence works as two exponential functions that influences the beam profiles. Our work is beneficial for the understanding the role of the interaction between spatial coherence and optical nonlinearity in the propagation of self-accelerating beams, which provides a new way to design reliable self-accelerating beams in nonlinear media. Our results thus can promote for further applications in optical manipulation and optical communication of complex environment.

\section{Theory}

In theory, the electric field of a typical coherent Airy beam at the source plane can be described as\cite{RN22}  
\begin{eqnarray}
    E\left({x},{y},z=0\right)={E}_{0}\text{Ai}\left(\frac{x}{{x}_{0}}\right)\exp\left(a\frac{x}{{x}_{0}}\right) \times \text{Ai}\left(\frac{y}{{x}_{0}}\right)\exp\left(a\frac{y}{{x}_{0}}\right)
\label{eq1}
\end{eqnarray}
where $Ai(x)$ is the Airy function, $a$ is a truncation factor, ${x}_{0}$ is a scale factor called characteristic length, and ${E}_{0}$ is a constant controlling the amplitude of Airy beam. Enlightened by previous studies\cite{RN15,RN29}, we introduce a random transverse spatial shifts $\Delta \Vec{{r}_{n}}$ to form an independent coherent mode $E_{n}(\Vec{r})=E(\Vec{r}+\Delta \Vec{{r}_{n}})$ at $z=0$. In fact, PIABs can be superposed by coherent modes $E_{n}(\Vec{r})$\cite{RN1,RN15,RN29}. Therefore, the coherent function $W\left(\Vec{{r}_{1}},\vec{{r}_{2}}\right)$ of PIABs can be defined as
\begin{eqnarray}
    W\left(\Vec{{r}_{1}},\vec{{r}_{2}}\right)=\underset{n}{\sum} {a}_{n}{E}^{*}_{n}(\Vec{{r}_{1}}) {E}_{n}(\vec{{r}_{2}})
\label{eq2}
\end{eqnarray}
 where ${a}_{n}$ are real values representing the power associated with coherent mode $E_{n}(\Vec{r})$. $\vec{{r}_{1}}$ and $\vec{{r}_{2}}$ represent the position vectors of two points in space, respectively. Clearly, a fully coherent beam is comprised of a single spatial mode, while a partially incoherent beam is comprised of an arbitrary number of modes. According to eq. (\ref{eq2}) and $I(\Vec{r})=\underset{n}{\sum}{a}_{n}{\lvert E(\Vec{r})\rvert}^2$, the spatial coherence function is $\mu(\Vec{{r}_{1}},\Vec{{r}_{2}})=W\left(\Vec{{r}_{1}},\vec{{r}_{2}}\right)/\sqrt{I(\vec{{r}_{1}})I(\vec{{r}_{2}})}$. Then, the coherence length $\delta$ can be defined as the transversal distance $\Delta r=\Vec{{r}_{1}}-\Vec{{r}_{2}}$ at which the spatial coherence function $\mu(\Vec{{r}_{1}},\Vec{{r}_{2}})$ drops to a certain fraction (e.g., $1/\text{e}$ or $1/{\text{e}}^{2}$) of its maximum value\cite{RN30}. For arbitrary coherent modes ${E}_{n}(\Vec{r})$, they all obey the paraxial wave equation by using slowly varying envelope approximation
\begin{equation}
    \frac{\partial {E}_{n}(\Vec{r})}{\partial z}=\frac{i}{2{k}_{0}n}\left(\frac{{\partial}^{2}{E}_{n}(\Vec{r})}{{\partial {x}^{2}}}+\frac{{\partial}^{2}{E}_{n}(\Vec{r})}{\partial {y}^{2}}\right)+i{k}_{0}\Delta n(I){E}_{n}(\Vec{r})
    \label{eq3}
\end{equation}
where ${k}_{0}$ is the vacuum wave vector, and $n$ is the refractive index. $\Delta n(I)$ is the nonlinear refractive index varying with $I$, where $I$ is the time average intensity of the beam. Based on eq. (\ref{eq2}) and eq. (\ref{eq3}), one can simulate the propagation of the PIAB in linear or nonlinear media.
\section{Experiment}

\begin{figure*}[ht]
\centering
	\includegraphics[scale=0.5]{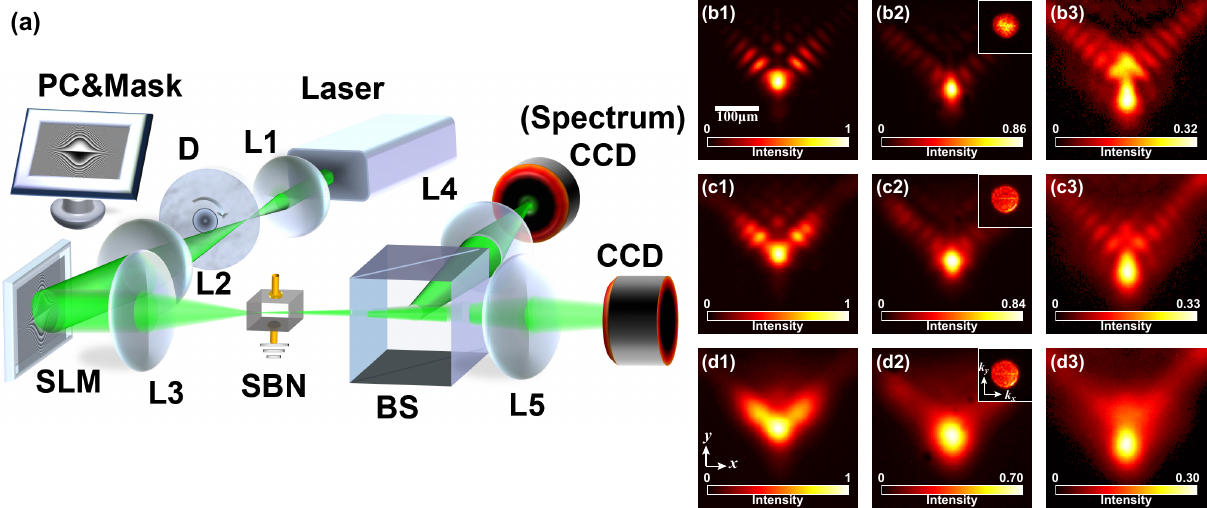}
	\caption{\label{fig. 1}(a) Schematic of experimental setup. SLM, spatial light modulator; SBN, strontium–barium–niobate crystal; BS, beam splitter; L, lens; D, rotating ground-glass disk; Mask, the cubic-phase pattern. (b)-(d) The experimental results of the PIABs with different coherence lengths without applying bias field on the SBN. The ${1}^{st}$ , ${2}^{nd}$ and the ${3}^{rd}$ columns are the input intensity pattern, output intensity pattern after 10 mm through crystal and output intensity pattern after 15 mm through crystal and air, respectively. The coherence length of (b)-(d) is infinity, $\delta =48$ $\upmu$m and $\delta =32$ $\upmu$m, respectively. The insets in (b2)-(d2) show corresponding Fourier spectra which are obtained by Fourier transform with lens L4.}
\end{figure*}

Our experimental setup for generation and propagation of PIAB is shown as Fig. \ref{fig. 1}(a): a linearly polarized Gaussian beam ($\lambda$=532 nm) transmits  through a rotating ground glass disk, such that a spatially incoherent light is generated \cite{RN1,RN15}. The partially incoherent light is then launched onto a spatial light modulator (SLM, Holoeye Pluto-VIS), where a cubic-phase pattern is loaded on the wave front. To produce a PIAB, a Fourier transform lens ${L}_{3}(f=150$ mm) is placed in front of the SLM, and the PIAB is obtained at the front focal plane of ${L}_{3}$. Here, the coherence length $\delta$ of the PIAB can be controlled by adjusting the diameter of the laser beam incident on the rotating diffuser \cite{RN26}. The PIAB is then launched into a biased 1-cm-long photorefractive media strontium-barium niobate (SBN:60) crystal for investigating its nonlinear propagation. By switching the polarity of the bias field ${E}_{0}$, self-focusing and self-defocusing nonlinearity can be achieved \cite{RN21}. The PIAB and its Fourier spectrum are monitored by two CCD cameras. In all experiments, the characteristic length ${x}_{0}$ and the truncation factor $a$ of PIABs are adopted as 14 $\upmu$m and 0.08, respectively. 

\section{Results and Discussion}
\subsection{Linear propagation of PIABs}
As for comparison, we first studied the linear propagation properties of the PIABs with different spatial coherence lengths. Here, there is no bias field on the SBN such that $\Delta n=0$, and the results of PIAB are shown in Figs. \ref{fig. 1}(b1)-\ref{fig. 1}(d3).  Apparently, as the coherence length decreases, the main lobe of the resulting Airy beam becomes larger with less side lobes, suggesting the limited coherence length working as an effective truncation factor $a$ of the Airy beam as shown in eq. (\ref{eq1}). In other words, even though we set the truncation factor as $a=0.08$ in all cases, the spatial incoherence enhances the truncating influence, which can be formulated as an additional exponential function $\exp(bx)$ with $b$ is a negetive value. For the PIABs, the fine structures of Airy beam are blurred as the coherence length decreases, and finally the beam profile becomes close to that of a Gaussian-like beam if the beam is fully incoherent. Nonetheless, the generated PIABs keep the self-accelerating property, and the transverse positions of the main lobes are shifted along the propagation direction as shown in Figs. \ref{fig. 1}(b1)-\ref{fig. 1}(d3) . Besides, the PIABs exhibit nearly the same transverse self-acceleration in the three cases, implying that the accelerating property of the PIAB weakly depends on the spatial coherence. However, the peak intensity changes during propagation look different. After comparing the results of the three columns , we find that the peak intensity decreases more with a shorter spatial coherence length. Therefore, the PIAB is more difficult to keep its non-diffracting features and the intensity decay is larger. 

\begin{figure}[h!]
\centering
	\includegraphics[scale=0.85]{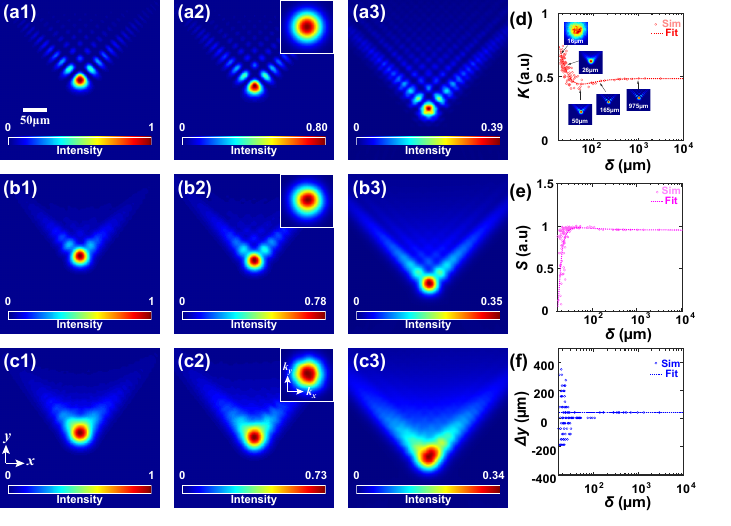}
	\caption{\label{fig:2} (a)-(c) The simulation results correspond to Figs. \ref{fig. 1}(b)-\ref{fig. 1}(d)  and  (d)-(f) the changes of the intensity ratio $K$, intensity profile similarity $S$ and transverse shifting of main lobe $\Delta y$ with different coherence lengths. The insets in (d) show the initial intensity of PIAB with coherence length 16 $\upmu$m, 26 $\upmu$m, 50 $\upmu$m, 165 $\upmu$m, 975 $\upmu$m, respectively.}
\end{figure}

To further prove above results, we also employed eq. (\ref{eq3}) to do the numerical simulation via multiple-phase-screen methods\cite{RN1}, and the results are as presented in Fig. \ref{fig:2}. Clearly, the numerical results agree well with the experimental result, demonstrating that PIABs still preserve the self-accelerating properties and the intensity profiles in Fourier space keep unchanged.

To quantitatively analyze the change of the diffracting property of the beams with or without the bias field, we define two factors for simple discussion: one factor is called peak intensity ratio $K={I}_{out}^{'}/{I}_{ini}$, where ${I}_{out}^{'}$ and ${I}_{ini}$ refer to the final and initial peak intensity of PIABs with nonlinear propagation, respectively; Another factor is named intensity profile similarity, $S=cov({I}_{out}^{'},{I}_{ini})/\sqrt{Var({I}_{out}^{'})Var({I}_{ini})}$, where $cov(X,Y)$ is the covariance between random distribution $X$ and $Y$, $Var(X)$ is the variance of random distribution $X$, ${I}_{out}^{'}$ and ${I}_{ini}$ refer to the final and initial intensity distributions of PIABs with nonlinear propagation, respectively. To quantitatively analyze the influence of the self-accelerating properties of PIABs after nonlinear propagation, we define the transverse shift of the main lobe $\Delta y$, which is the spatial shift of the main lobe after 5 mm through air.

In the linear case, the calculated results of $K$ and $S$ with different coherence lengths are plotted in Figs. \ref{fig:2}(d)-\ref{fig:2}(e). When coherence length is larger than 30 $\upmu$m, the decrease of incoherence does not affect beams structure of Airy beams much. The peak intensity ratio $K$ is almost unchanged (first goes down little until around 50 $\upmu$m and then goes up), as well as the intensity profile similarity $S$. PIABs possess a stable transverse accelerating, as presented in Fig. \ref{fig:2}(f). However, when coherence length is smaller than 30 $\upmu$m, $K$ quickly increases with a decreasing coherence length while $S$ decreases. This behavior is same as that by increasing  the truncation factor $a$ for a fully coherent Airy beam. Especially, when the coherence length is very small (16 $\upmu$m), the intensity distributes randomly (the $1^{st}$ inset of Fig. \ref{fig:2}(d)) and $S$ is nearly zero though $K$ is close to one, as a result of the strong distortions of the incoherence, i.e., a large effective truncation factor $a$. Beams have evolved into Gaussian-like beams and the diffracting is too strong. PIABs cannot keep the properties of self-accelerating and close-nondiffracting. By employing an additional exponential function to the wavefunction of a fully coherent Airy beam, the simulation results can predict the results for PIABs, confirming our theoretical assumption that spatial incoherence can be formulated as an effective exponential function.

\subsection{Nonlinear propagation of PIABs}

\begin{figure*}[htbp!]
\centering
	\includegraphics[scale=0.85]{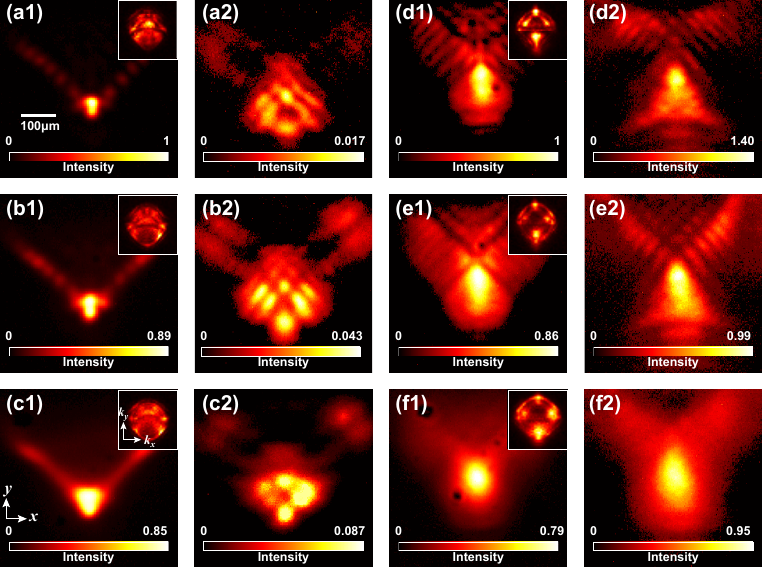}
	\caption{\label{fig:3} The output intensity patterns of PIABs with different spatial coherence after 10 mm through crystal ($1^{st}$ column and $3^{rd}$  column) plus another 5mm through air ($2^{nd}$ column and $4^{th}$ column). The insets in the upper right corner in (a1)-(f1) show corresponding Fourier spectra. (a)-(c) self-focusing nonlinearity (${E}_{0}=3\times 10^4$ $\text{V/m}$), (d)-(f) self-defocusing nonlinearity (${E}_{0}=-3\times 10^4$ $\text{V/m}$). $1^{st}$, $2^{nd}$ and $3^{rd}$ rows correspond to $\delta =\infty$ ,$\delta=48$ $\upmu$m and $\delta =32$ $\upmu$m, respectively.}
\end{figure*}

Next, we experimentally study the nonlinear propagation of the PIABs in a biased SBN:60 crystal. The saturable nonlinearity of SBN for an e-polarized beam can be determined by $\Delta n=-0.5{n}_{e}^{3}{\gamma}_{33}{E}_{0}/(1+I)$, in which ${n}_{e}=2.3$ is the unperturbed refractive index, ${\gamma}_{33}=280$ $\text{pm/V}$ is the nonlinear coefficient, ${E}_{0}$ is the amplitude of the bias field and $I$ is the time average intensity of the beam. When a positive bias field of ${E}_{0}=3\times 10^4$ $\text{V/m}$ is applied, the PIABs experience a self-focusing nonlinearity. As a result, the lobes of the beams with different spatial coherence all self-trap into smaller lobes [Figs. \ref{fig:3}(a1)-\ref{fig:3}(c1)]. To further observe how the incoherence affects propagation property of PIABs, we let these self-focused beams propagate in a subsequent free-space (5-mm long air). The final results are shown in Figs. \ref{fig:3}(a2)-\ref{fig:3}(c2). For the fully coherent case, the shape of Airy beam is strongly deformed and peak intensity cannot keep at in the main lobe (Airy “head”) though the acceleration of main lobe still exists [Fig. \ref{fig:3}(a2)]. Its Fourier spectrum in \textbf{k}-space is partially focused at the center shown in the inset of Fig. \ref{fig:3}(a1), suggesting that the Airy beam exhibits normal diffraction. In contrast, a PIAB can keep somewhat Airy-like pattern [Figs. \ref{fig:3}(b2)-\ref{fig:3}(c2)]. The peak intensity not only stays at the main lobe, but also maintains the obvious self-acceleration. Modulated by more incoherence, PIAB driven by a self-focusing nonlinearity keep the shape and self-accelerating propagation properties better. Moreover, PIAB with more incoherence has a larger output peak intensity and intensity ratio $K$ after the subsequent 5-mm propagation in free space. Less intensity in \textbf{k}-space is found to be focused toward the center [the insets of Figs. \ref{fig:3}(b1)-\ref{fig:3}(c1)], verifying that spatial incoherence reduces the degradation of the beam quality caused by self-focusing nonlinearity.

It is known that when self-focusing nonlinearity is employed, it would firstly balance the diffraction of the beams and then break down the original interference of the plane waves that contribute to the nondiffraction and self-acceleration of Airy beams\cite{RN27,RN28}. Thus, the shape of an Airy beam is deformed and the beam cannot keep self-acceleration or a good intensity output. However, if the diffraction of Airy beam becomes stronger, self-focusing nonlinearity would be difficult to break down the beam structure and propagation properties of Airy beams\cite{RN21}. Diffraction of a PIAB is determined by its coherence length while diffraction of coherent case depends on its beam diameter. So, the more incoherent, the more difficult it is for self-focusing nonlinearity to break down the beam structure and propagation properties of Airy beams. Breaking down the shape of a PIAB requires stronger self-focusing nonlinearity to balance the stronger diffraction. Clearly, appropriate spatial incoherence can resist the corruption of beam quality caused by self-focusing nonlinearity.

By changing the polarity of the bias field(${E}_{0}=-3\times 10^4$ $\text{V/m}$), PIABs experience a self-defocusing nonlinearity. In this case, the shapes of all Airy beams spread much more but the peak intensity after subsequent linear propagation in the air decreases less as compared to the linear case [Figs. \ref{fig:3}(d)-\ref{fig:3}(f)] while the beams keep a good transversal acceleration. Especially, the Fourier spectrum in all the cases reshapes into a diamond-like pattern in \textbf{k}-space, as shown in the insets of Fig. \ref{fig:3}(d1)-\ref{fig:3}(f1), resembling the first Brillouin zone of an asymmetric square lattice as in Ref. \cite{RN21}. It implies that PIABs will experience an anomalous diffraction when driven by a self-defocusing nonlinearity and their output peak intensity is stronger than input peak intensity during subsequent linear propagation. However, such effect gets weaker for the beams of more incoherent, and the beams can maintain the Airy-like pattern better after the subsequent linear propagation. Clearly, increasing the spatial incoherence indeed resist the degradation of beam quality caused by self-defocusing nonlinearity.

\begin{figure}[htbp!]
\centering
	\includegraphics[width=8.6cm]{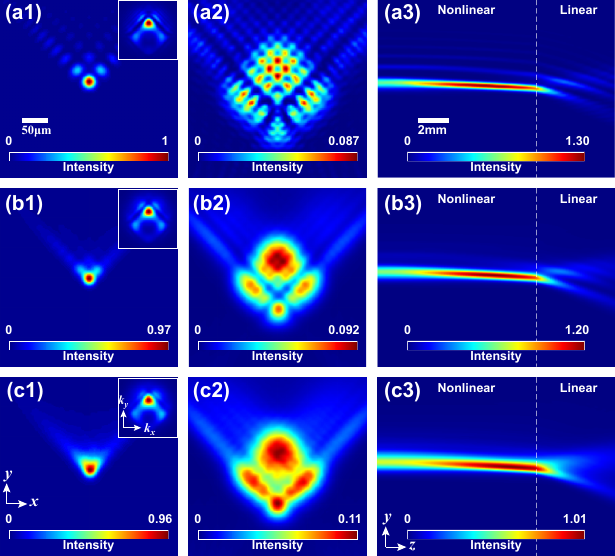}
	\caption{\label{fig:4} Numerical simulation of propagation of PIABs under an initial self-focusing nonlinearity (${E}_{0}=3\times10^4$ $\text{V/m}$). (a1)-(c1) and (a2)-(c2) correspond to the experimantal reuslts in Figs. \ref{fig:3}(a1)-\ref{fig:3}(c1)  and  \ref{fig:3}(a2)-\ref{fig:3}(c2), respectively. (a3)-(c3) are the corresponding sideviews.}
\end{figure}

\begin{figure}[htbp!]
\centering
	\includegraphics[width=8.6cm]{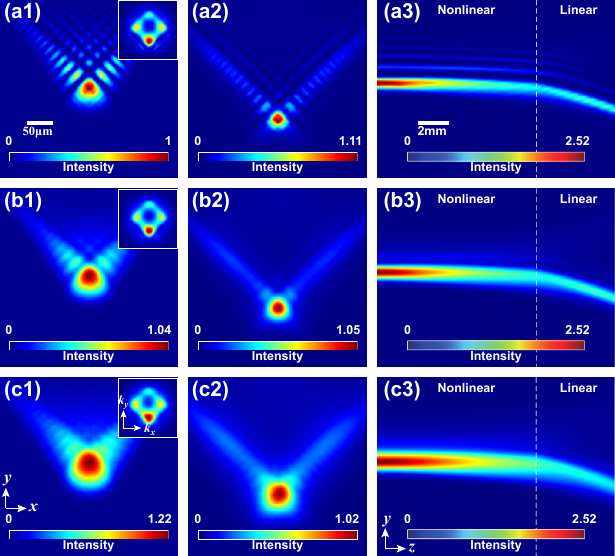}
	\caption{\label{fig:5} Numerical simulation of PIABs propagation under an initial self-defocusing nonlinearity(${E}_{0}=-3\times 10^4$ $\text{V/m}$). (a1)-(c2) correspond to Figs. \ref{fig:3}(d)-\ref{fig:3}(f). (a3)-(c3) are the corresponding sideviews.}
\end{figure}

To further confirm our experimental observations, we did simulations with the same parameters as those in experiment. As shown in Fig. \ref{fig:4}, when ${E}_{0}=3\times 10^4$ $\text{V/m}$ is applied, numerical results are highly consistent with the experimental results. PIABs driven by a self-focusing nonlinearity keep their shapes better and self-accelerating propagation with a stronger spatial incoherence. In addition, less intensity in \textbf{k}-space is as expected to focused onto the center, further verifying that spatial incoherence reduces the self-focusing nonlinear effect. Moreover, from Figs. \ref{fig:4}(a3)-\ref{fig:4}(c3), one can find that incoherent cases always keep the transverse peak intensity stay at main lobe longer and the longitudinal peak intensity along propagation directions appears latter, indicating that spatial coherence length can control the transversal and longitudinal positions of peak intensity. With a reversed bias field of ${E}_{0}=-3\times 10^4$ $\text{V/m}$, self-defocusing exhibits a weaker corruption of the beam structures of Airy beams and the beams maintain their patterns and self-accelerating well, as shown in Fig. \ref{fig:5}. Furthermore, as the spatial coherence length decreases, the PIABs keep the pattern better after the subsequent free-space propagation, exhibiting a better propagation effect compared with the linear case. These results can be also seen clearly from the sideviews of beam propagation [Figs. \ref{fig:4}(a3)-\ref{fig:4}(c3) and \ref{fig:5}(a3)-\ref{fig:5}(b3)].

\begin{figure*}[htbp!]
\centering
	\includegraphics[width=12cm]{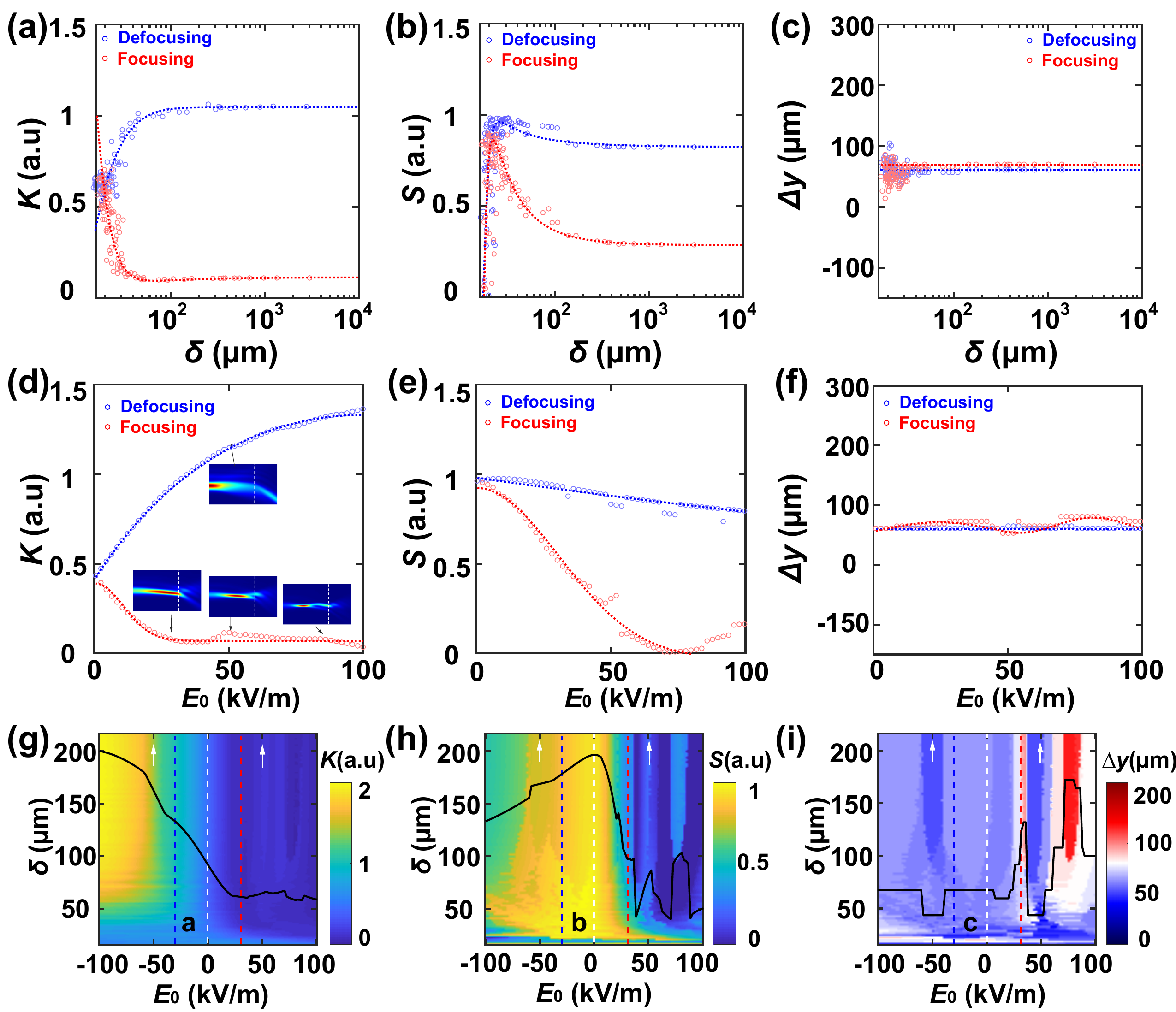}
	\caption{\label{fig:6} Changes of peak intensity ratio $K$, intensity profile similarity $S$ and transverse shifting of main lobe $\Delta y$ of PIABs with different conditions: (a)-(b) different coherence length with the same initial nonlinearity ($|{E}_{0}| =300\times 10^4$ $\text{V/m}$); (c)-(d) different nonlinearity with $\delta=48$ $\upmu$m; (g)-(i) different spatial incoherence and driven by different initial nonlinearity. Black curves indicate the changes of $K$, $S$, and $\Delta y$ with ${E}_{0}$ at $\delta=200$ $\upmu$m.}
\end{figure*}

Similar to the linear case, we also observe the quantitative variation of the intensity ratio $K$, intensity profile similarity $S$ and transverse shifting of main lobe $\Delta y$ of main lobe under different coherence and nonlinearity [Fig. \ref{fig:6}]. When the coherence length increases, intensity ratio $K$ in a fixed self-focusing case (${E}_{0}=3\times 10^4$ $\text{V/m}$) shows a sharper and larger exponential fall due to the further attenuation of  breaking down of beam structure by the self-focusing effect [Fig. \ref{fig:6}(a)]. Meanwhile, as a result of the same self-focusing effect, the change of the intensity profile similarity $S$  is much different from that of the linear case [Fig. \ref{fig:6}(b)]. At the beginning, $S$ still keeps going up with coherence length. However, from 26 $\upmu$m, $S$ does not preserve as linear case. It will quickly declines from 0.9 to 0.3 until 200 $\upmu$m and then keep almost unchanged. For a fixed self-defocusing case (${E}_{0}=-3\times 10^4$ $\text{V/m}$), a similar but much less obvious phenomenon happens in intensity profile similarity $S$ . However, due to the enhancement of self-defocusing on the output intensity (anomalous diffraction) shown in Fig. \ref{fig:6}(a), the intensity ratio $K$ experience a quick rise and turns into a stable state after 70 $\upmu$m. For both of the two nonlinear cases, the transverse accelerating shift maintains stable while coherence varies [Fig. \ref{fig:6}(c)]. Generally, self-focusing with ${E}_{0}=3\times 10^4$ $\text{V/m}$ enhances the accelerating while self-defocusing does not affect it unless the incoherence is strong.

In addition, for the self-focusing nonlinearity, when the coherence length is fixed to 48 $\upmu$m ($>30$ $\upmu$m), which means the truncating effect of incoherence is still weak, the intensity ratio $K$ and $S$ sharply go down with the increasing biased field and will have an obvious jumping at a certain value (here, $\sim 5\times 10^4$ $\text{V/m}$) [Figs. \ref{fig:6}(d)-\ref{fig:6}(e)]. This jump is caused by the soliton shedding of  Airy beams\cite{RN21} generated by strong self-focusing nonlinearity, i.e., there is a new self-focusing Airy soliton beginning to form at this jumping biased field, as shown in the insets in Fig. \ref{fig:6}(d). Interestingly, while the biased field is larger than $7.5\times 10^4$ $\text{V/m}$, $S$ goes up, implying that the beam pattern changes back to an Airy profile. As mentioned in Ref.\cite{RN21}, this is possible to be a result of soliton shedding and the self-healing property of Airy beams. Note that, the transverse accelerating shift during subsequent linear propagation in free space also changes with the periodic Airy solitons and shows snake oscillation [Fig. \ref{fig:6}(f)]. Contrary to that, for the self-defocusing case, no jumping or snake oscillation happens. As the biased field increases, $K$ keeps increasing and can be larger than one [Fig. \ref{fig:6}(d)], indicating the anomalous diffraction is getting stronger with an increasing self-defocusing nonlinearity. However, the increase of $K$ is getting smaller and smaller and $K$ may reach a saturated state finally. Meanwhile, $S$ decreases slowly and the transverse accelerating shift has no change, demonstrating the Airy-like structure and profiles of PIABs keep good, and self-defocusing nonlinearity produces a very weak influence. Definitely, this influence gets stronger with a bigger biased field.

To further investigate the combined interaction of incoherence and nonlinearity, we plotted the detailed results of PIABs with different coherence and nonlinearity in Figs. \ref{fig:6}(g)-\ref{fig:6}(i). From Figs. \ref{fig:6}(g)-\ref{fig:6}(h), one can see that the larger coherence or biased field can enhance the anomalous diffraction caused by self-defocusing, leading to an increasing intensity ratio $K$ and intensity similarity $S$. However, for the focusing case, more coherence or nonlinearity causes more serious corruption of propagation properties and beam structure, resulting in smaller $K$ and $S$. Especially, the biased field inducing jumping decreases with a larger coherence length (See the black lines in Fig. \ref{fig:6}). In this case, $K$, $S$, and $\Delta y$ present quicker and stronger periodic fluctuations though $K$ and $S$ still keep a decreasing trend while $\Delta y$ keeps an increasing trend. Interestingly, for self-defocusing nonlinearity, the decline of $S$ and enhancement of $K$ also get stronger and stronger while incoherence becomes weaker, and $\Delta y$ exhibit a jumping at coherence lengths larger than around 60 $\upmu$m. Definitely, when coherence length is smaller than 30 $\upmu$m, the PIABs are seriously broken down by the random distortion of huge incoherence. In general, PIABs with proper spatial incoherence always have the better robustness for nonlinear effect.

\section{Conclusion}
In summary, we have quantitatively investigated the nonlinear propagation of PIABs both in theory and experiment. It is found that, the influence of coherence and nonlinearity on the propagation of PIABs can be formulated as two effective exponential functions. Moreover, spatial incoherence and nonlinearity have opposite effects on the propagation properties of PIABs, and thus their interaction can be used to improve beam quality of Airy beam. By controlling the spatial incoherence, PIABs can resist the corruption on the beam quality caused by nonlinear propagation. Under a appropriate balance between incoherence and nonlinearity, PIABs present much better beam quality after a long propagation distance. Our results provide a comprehensive understanding on the nonlinear propagation dynamics of PIABs, beneficial to the development and control of Airy beam and the corresponding applications.

\begin{backmatter}
\bmsection{Funding}
the National Natural Science Foundation of China (11604058), the Guangxi Natural Science Foundation (2020GXNSFAA297041, 2020GXNSFDA238004, 2016GXNSFBA380244), Innovation Project of Guangxi Graduate Education(YCSW2022041, YCSW2023011), Sichuan Science and Technology Program (2023NSFSC0460), and Open Project Funding of the Ministry of Education Key Laboratory of Weak-Light Nonlinear Photonics (OS22-1).

\bmsection{Acknowledgments}

\bmsection{Disclosures}
The authors declare that there are no conflicts of interest related to this article.

\bmsection{Data Availability}
Data underlying the results presented in this paper are not publicly available at this time but may be obtained from the authors upon reasonable request.
\end{backmatter}


\end{document}